\documentclass{aa}
\usepackage{graphicx}
\begin{document}
   \title{Time-resolved, multi-color photometry and spectroscopy of Virgo 4 (OU Vir): a high orbital inclination, short orbital period dwarf nova}

   \subtitle{}

   \author{E. Mason
          \inst{1}
          \and
          S. B. Howell\inst{2}
	\and  
	P. Szkody\inst{3}
	\and  
	T. E. Harrison\inst{4}
	\and
	  J. A. Holtzman \inst{4}
         \and D. W. Hoard \inst{3}
	    }

   \offprints{E. Mason}

   \institute{ESO, Alonso de Cordova 3017, Casilla 19001, Vitacura, 
		Santiago, Chile \\
              \email{emason@eso.org}
         \and
             Astrophysics Group, Planetary Science Institute, Tucson, AZ 85705, USA \\
             \email{howell@psi.edu}
	\and
	     Department of Astronomy, University of Washington, Seattle, WA 98195, USA \\
	     \email{szkody@astro.washington.edu hoard@astro.washington.edu}
	\and
	     Department of Astronomy, New Mexico State University, Las Cruces, NM 88003, USA \\
	     \email{tharriso@nmsu.edu holtz@nmsu.edu}
           	       }

   \date{Received Jul, 2002; accepted ?, ?}

   \abstract{
   We present multi-color photometry and time resolved spectroscopy of OU~Vir. The analysis of the quiescent light curve shows that OU~Vir is characterized by  i) strong cycle-to-cycle brightness variations, and ii) hot spot modulated light curve with grazing eclipse of the impact region. Colors are derived both in- and out- of eclipse. The time-resolved spectroscopy allows us to produce the radial velocity curve from the H$\alpha$ accretion disk emission line which possibly reveals only weak evidence for hot spot line emission. The hot spot is believed to be a turbulent optically thick region, producing mostly continuum emission.
   }

   \titlerunning{OU~Vir}
   \authorrunning{E. Mason et al.}
   \maketitle
%

\section{Introduction}

   The object LBQS 1432-0033 (alias Vir~4 in Downes et al. 1997 catalog, OU ~Vir in Downes et al. 2001 catalog) was discovered by Berg et al. (1992) as 
part of their quasi-stellar object spectroscopic survey. Downes et al. (1997) 
list the object as a cataclysmic variable (CV) of unknown type 
with B$_{pg}$=18.5 based on the Berg et al. spectrum. Vanmunster et al. (2000) 
present a photometric review for the object indicating that OU~Vir is an 
 eclipsing CV with a period of 1.75 hr and has been seen in outburst and 
probably in superoutburst. These authors quote an outburst amplitude of near 
4 magnitudes with a probable inter-outburst period near 79 days. The
only other published spectrum of OU~Vir is in Szkody et al. (2002).

The spectra by Berg et al. (1992) and Szkody et al. (2002), are characterized by strong double peaked emission lines from the accretion disk. Eclipses and double peaked lines are both signatures of high orbital inclination systems. 
High orbital inclination systems are preferred targets for the understanding of the binary system geometry and the accretion process, due to the large amount of information provided by both their light curve and time resolved spectroscopy. 
 
Here, we present new  photometric and spectroscopic observations for 
OU~Vir. Our multi-color photometry reveals the stars behavior in the B through I bands. Our time-resolved spectroscopic observations of 
OU~Vir have been used to produce the first radial velocity curve for this 
object, as well as to study the spectrum throughout the orbit. 
We discuss the data collection and reduction in 
Sect.~2, our analysis in Sect.~3, and a discussion of our
results in Sect. 4.

\section{Observations}

\subsection{Multi-color Photometry}

Time resolved multi-color photometry of OU~Vir was obtained using the
New Mexico State University 1 m telescope located at Apache Point 
Observatory\footnote{http://loki.apo.nmsu.edu/$\sim$tcomm/index.html}.
OU~Vir was observed over eight nights between 2001 April 13 and May 1 using
an Apogee 512x512 thinned back-side illuminated CCD with standard 
Johnson/Cousins BVRI filters. On each of the nights, one entire orbital period
was covered by cycling through the four filters. Exposure times were three
minutes in B, and two minutes in V, R, and I. To increase the precision of 
the timing of the photometric eclipse, OU~Vir was observed for a complete cycle on May 1 using only the V-band filter. The log of the photometric observations can be found in Table~1. 

\begin{table*}[H]
\begin{center}
\scriptsize
\caption{\small Log of observation of the photometric run.}
\begin{tabular}{ccccc}
 & & & & \\
UT date &\multicolumn{4}{c}{UT-start/UT-end, points} \\
 & & & & \\
 & B & V & R & I \\
13 April 2001 & 05:30:57-07:30:17, 11 & 05:34:51-07:33:09, 13 & 05:37:15-07:35:34, 12 & 05:39:38-07:37:56, 12 \\
23 April 2001 & 05:13:55-07:13:26, 12 & 05:17:48-07:16:20, 12 & 05:20:12-07:18:43, 12 & 05:22:36-07:19:07, 12 \\
24 April 2001 & 05:02:36-06:51:50, 11 & 05:06:29-06:53:44, 11 & 04:58:18-06:57:09, 12 & 05:00:42-07:27:54, 13 \\
25 April 2001 & 05:20:39-07:20:13, 12 & 05:24:33-07:23:07, 12 & 05:26:57-07:25:31, 12 & 05:30:20-07:26:54, 11 \\
27 April 2001 & 05:30:41-07:30:14, 12 & 05:34:35-07:33:09, 12 & 05:36:59-07:35:35, 12 & 05:39:21-07:35:57, 12 \\
28 April 2001 & 05:11:11-07:11:11, 12 & 05:15:12-07:14:08, 12 & 05:17:40-07:16:34, 12 & 05:20:10-07:18:58, 12 \\
29 April 2001 & 05:29:11-07:29:16, 12 & 05:33:04-07:32:12, 12 & 05:35:28-07:34:36, 12 & 05:37:51-07:37:04,12 \\
30 April 2001 & 08:09:15-10:06:21, 12 & 08:11:07-10:09:15, 12 & 08:13:30-09:50:29, 10 & 08:15:53-10:14:03, 12 \\
1 May 2001 & -  & 08:39:33-10:41:54, 55 & - & - \\
\end{tabular}
\end{center}
\end{table*}

Data reduction was carried out according to standard routines, and aperture 
photometry was performed on OU~Vir and four other nearby field stars in order to obtain differential magnitudes (e.g. Howell et al. 1988). On 2001 June 6, these field stars were calibrated to the system of Landolt by observations of the standard stars SA102-58, SA102-620, SA105-66, SA106-485, SA107-544. 

\subsection{Spectroscopic Observations}

\begin{table}[b]
\begin{center}
\scriptsize
\caption{\small Log of observation of the spectroscopic run. Due to the time in the headers being inconsistent with respect to the telescope console time, we have recorded the latter one in this table. It is recorded only to the minute.}
\begin{tabular}{ccc}
& & \\
 UT Start-Exposure & Wavelength & Integration Time (sec) \\
& & \\
08:27 & Red & 300 \\
08:38 & Red & 600 \\
08:49 & Red & 600 \\
08:59 & Red & 600 \\
09:12 & Red & 600 \\
09:22 & Red & 600 \\
09:33 & Red & 600 \\
09:46 & Red & 600 \\
09:50 & Red & 600 \\
10:07 & Red & 600 \\
10:19 & Red & 600 \\
10:30 & Red & 600 \\
10:40 & Red & 600 \\
10:53 & Red & 600 \\
11:04 & Red & 600 \\
11:14 & Red & 600 \\
11:25 & Red & 600 \\
11:37 & Red & 600 \\
11:53 & Red & 600 \\
11:59 & Red & 600 \\
12:13 & Blue & 900  \\
\end{tabular}
\end{center}
\end{table}

OU~Vir was observed using the red channel spectrograph at the MMT Observatory on the night of 29 March 2001 UT. A single 900 sec blue (4000-5250\AA) spectrum
was obtained along with a time-series of 600 sec 
red (5600-8500\AA) exposures. 
Table 2 gives a log of the observations.
The setup used a 1 arcsec slit giving spectral resolutions of 0.5 \AA/pixel in the blue and 1.0 \AA/pixel in the red. The night appeared to be photometric and had constant seeing ($\sim$1.2 arcsec), as evidenced by examination of the slit viewer images. Brightness changes of OU~Vir were also quite obvious in the slit viewer monitor. 

The data were reduced in IRAF using average bias and red/blue flat fields taken at the start of the night. HeNeAr arc lamp exposures were obtained at each
standard star position before and after every three OU~Vir observations.
The standard star HZ 44, used to flux calibrate the OU~Vir spectra, was observed near in time and airmass to the blue spectrum and at the 
start of the red time-series sequence. The wavelength calibration had formal
errors of 0.1\AA \ across the wavelength range 
and the absolute flux calibration
appears good to near 15\% as estimated by using HZ 44 to flux calibrate  
another standard star.

\section{Data Analysis}

\begin{figure*}
\centering
\rotatebox{-90}{\includegraphics[width=12cm]{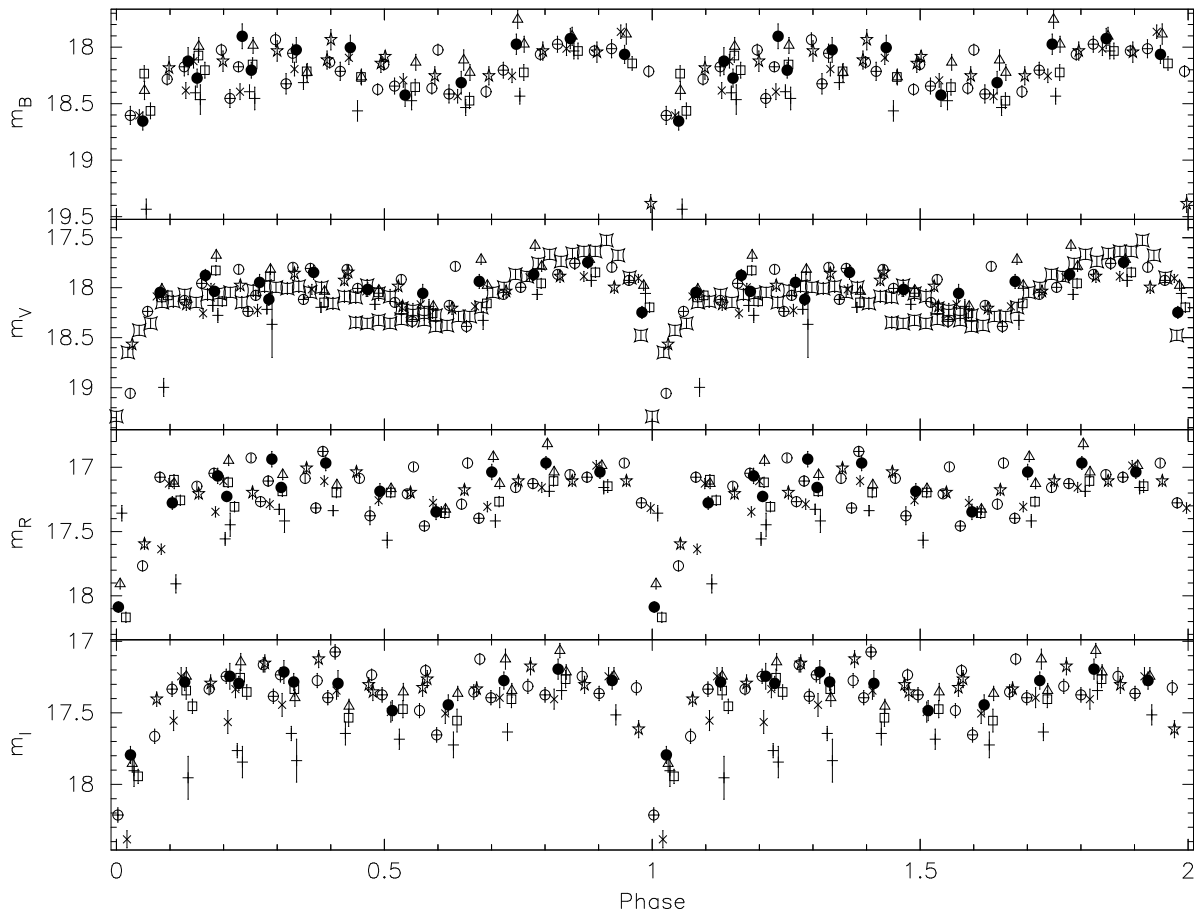}}
\caption{Multi-color photometry of OU~Vir. Different symbols refer to different nights of observations: -- 13/Apr, $\circ$ 23/Apr, $\times$ 24/Apr, ``boxes'' 25/Apr, $\triangle$ 27/Apr, $\oplus$ 28/Apr, $\bullet$ 29/Apr, $\star$ 30/Apr. The ``pillow shaped boxes'' refer to the V band photometry of 01/May. Data points have been plotted over two cycles to better view of the eclipse.}
\end{figure*}

Upon phasing our photometric and spectroscopic data we encountered the problem of choosing an appropriate ephemeris. Vanmunster et al. (2000) determined the OU~Vir orbital period and ephemeris by fitting six eclipses during the June 2000 outburst. Phasing of our photometric data according their ephemeris produced light curves showing eclipse minima at phase $\sim$0.8. The phase offset between our and Vanmunster et al. (2000) eclipse minima is easily explained by the error propagation of the uncertainties in their ephemeris. On the other hand, the time resolution of both our photometric and spectroscopic data is not high enough to produce accurate ephemeris.  
We thus, decided to phase all our data using the orbital period determined by Vanmunster et al. (2000)\footnote{We searched our radial velocity measurements for periodicity by using the PDM routine (Stellingwerf 1978) finding an orbital period of $\sim$102 min, in quite good agreement with the photometric observation by Vanmunster et al. (2000). However, the resulting uncertainties on the orbital period was of $\sim$20 min due to the relatively low number of spectra in hand.}, but we used the time $T_0$ corresponding to the eclipse minimum of our May 1$^{st}$ V band light curve (which however still has a time resolution of only 2-3 min). 

\subsection{Photometry} \label{s31}

We have plotted the multi-color light curves of OU~Vir in Fig. 1. In Fig.~1, each single night is plotted using a different symbol in order to show the 
strong intrinsic brightness variations of OU~Vir. 
Despite the night by night brightness variations we analyzed each waveband light curve using all data points at a time, as single night light curves have too few points. 

Averaging the points in each light curve (excluding the eclipse) we find the following mean colors: V = 18.08, B $-$ V = 0.14, V $-$ R =0.95, and V $-$ I =0.68. Rough eclipse colors have been computed averaging the two faintest data points in the run for each waveband. We found V$_e$=19.17, B-V $_e$= 0.24, V-R $_e$=1.05 and V-I $_e$=0.87. 
Eclipse colors are redder than out of eclipse, possibly providing evidence of secondary star contribution. 
 
\begin{table*}[H]
\begin{center}
\scriptsize
\caption{\small Photometric magnitudes determined from the BVRI light curves in Fig.~1. Each magnitude was derived through a weighted average of the data points in the phase range specified. Last column (ID) roughly identifies the measured feature in the light curve. ``average'' refers to the average magnitude computed excluding the in-eclipse data points.}
\begin{tabular}{ccccc}
 &  &  &  & \\
Color & Phase range & Instr. Mag. &  Used data points & ID \\
 &  &  &  & \\
B & 0.14-0.73 & 18.23$\pm$0.01 & 62 & average \\
B & 0.79-0.96 & 18.00$\pm$0.02 & 14 & pre-eclipse hump \\
B & 0.49-0.71 & 18.29$\pm$0.01 & 22 & secondary minimum\\
V & 0.10-0.71 & 18.08$\pm$0.01 & 99 & average \\
V & 0.81-0.94 & 17.76$\pm$0.01 & 17 & pre-eclipse hump \\
V & 0.52-0.70 & 18.16$\pm$0.01 & 32 & secondary minimum\\ 
R & 0.15-0.70 & 17.18$\pm$0.01 & 54 & average \\
R & 0.80-0.96 & 17.04$\pm$0.01 & 15 & pre-eclipse hump \\
R & 0.53-0.70 & 17.24$\pm$0.01 & 14 & secondary minimum\\
I & 0.17-0.88 & 17.35$\pm$0.01 & 72 & average \\
I & 0.76-0.88 & 17.28$\pm$0.02 & 11 & pre-eclipse hump \\
I & 0.51-0.65 & 17.42$\pm$0.02 & 14 & secondary minimum\\
 &  &  &  & \\
\end{tabular}
\end{center}
\end{table*}

Common features to each light curve are the deep eclipse and a pre-eclipse hump. The eclipse is about 1 mag deep and shows asymmetric ingress and egress profiles (the ingress being steeper than the egress), in each waveband. 
The pre-eclipse hump varies in amplitude depending on 
the photometric waveband. The pre-eclipse hump is particularly evident in the B and V bands ($\sim$0.3 mag and 0.22 mag, respectively), and decreases in amplitude at longer wavelengths. 
A closer inspection of the B and V light curves shows a 
weak secondary maximum near phase $\sim$0.3 and/or a secondary minimum at 
phase $\sim$0.6.  In Table~3, we list the average magnitude for each of the above mentioned features in each light curve. 
We believe that the light curve shape is due to the hot spot which is seen from the outside around orbital phase 0.9 (the pre-eclipse hump) and from the inside around phase $\sim$0.4. Similar double humped light curves have been
observed in the high inclination, short orbital period systems WZ~Sge 
(Skidmore 1998), OY~Car (Schoembs \& Hartmann 1983), Z~Cha (Wood et al. 1986, van Amerongen et al. 1990), and DV~UMa (Howell et al. 1988).  Causes for these types of modulations are either optically thin accretion disks with an anisotropically emitting hot spot, or an optically thick accretion disk which partially eclipses the hot spot when viewed from the inside. 

\begin{table}
\begin{center}
\scriptsize
\caption{Continuum and line flux values as measured in the Blue spectrum and the average red spectrum.}
\begin{tabular}{cccc}
Line & Flux & Continuum & Spectrum \\
(\AA) & (erg cm$^{-2}$ sec$^{-1}$ \AA$^{-1}$) &(erg cm$^{-2}$ sec$^{-1}$ \AA$^{-1}$) \\ 
& & & \\
4101.7 & 1.60E-14 & 8.99E-16 & Blue \\
4340.5 & 2.07E-14 & 7.59E-16 & Blue \\
4471.6 & 4.82E-15 & 6.84E-16 & Blue \\
4861.3 & 2.44E-14 & 5.75E-16 & Blue \\
6562.8 & 5.86E-14 & 8.24E-16 & Red \\
6678.1 & 3.81E-15 & 7.94E-16 & Red \\
7065.2 & 3.81E-15 & 7.00E-16 & Red \\
\end{tabular}
\end{center}
\end{table}

Average BVRI magnitudes and colors were also computed for each single night\footnote{In this case, no eclipse points were ruled out due to the poor time resolution which does not allow us to properly distinguish the features in each light curve.}. Plots of the night average magnitude vs. time showed that the intrinsic brightness variations are random and with no trend and/or long term variations. 
The colors vary by $\sim$0.1 mag and these small variations are not correlated with the magnitude. 

\subsection{Spectroscopy} \label{s32}

\begin{figure*}
\centering
\rotatebox{-90}{\includegraphics[width=12cm]{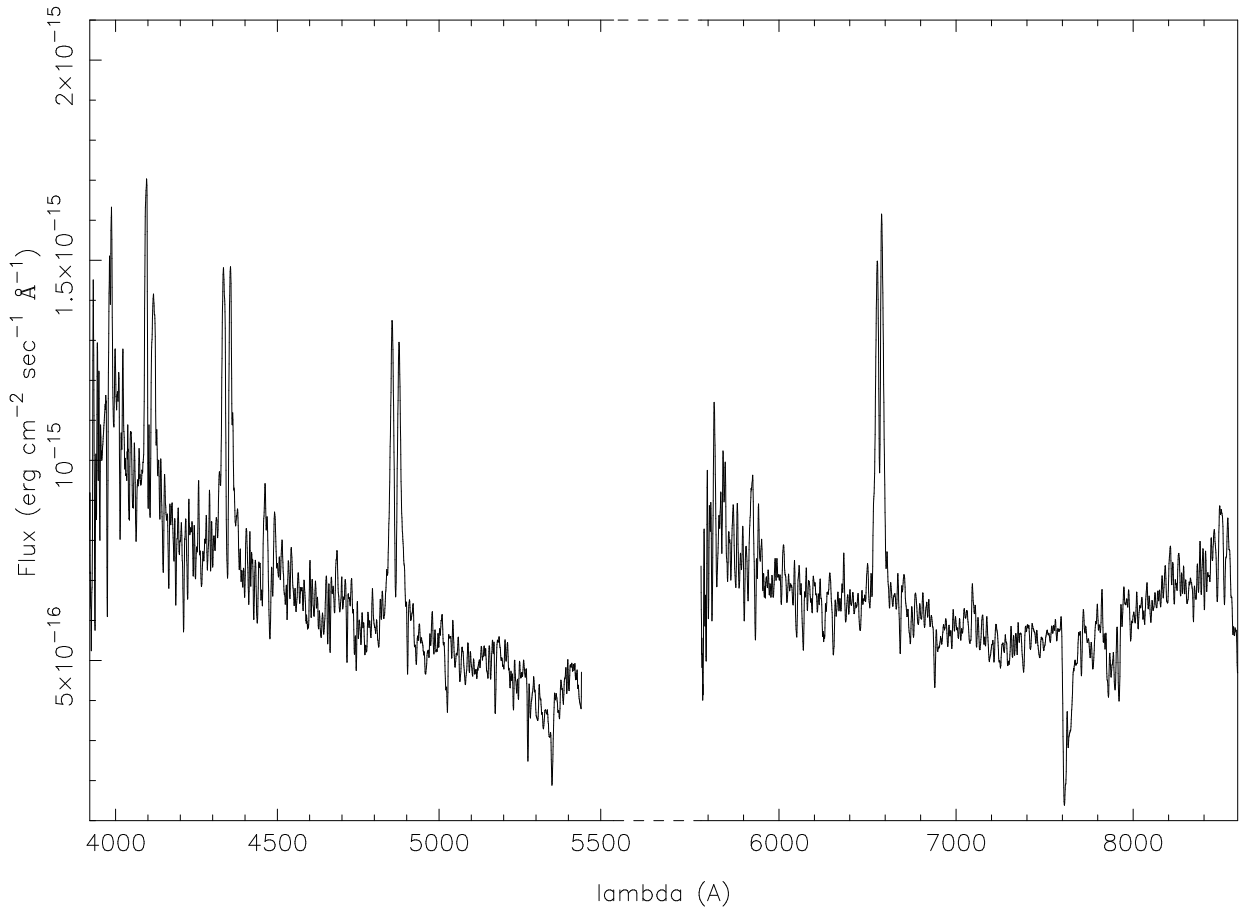}}
\caption{Blue and Red spectra of OU~Vir. The Blue spectrum is centered at orbital phase $\sim$0.78; the Red spectrum was taken at phase 0.79 of a previous orbital cycle. The discrepant continuum level is explained both by different slit losses in the two exposures, and by intrinsic brightness changes in the binary system (e.g. see Fig.~7). Both spectra have been smoothed with a box size of 5 pixels.}
\end{figure*}

The OU~Vir Red spectra show the emission lines H$\alpha$ and  HeI 
5876, 6678, and 7065, while the Blue spectrum shows the Balmer lines H$\delta$, H$\gamma$, and H$\beta$, the HeI lines 4471 and 5016, the Fe~II lines 5018 (blending with He~I 5016) and 5169,  and the HeII line 4686 (Fig.~2). Line flux measurements (for the strongest lines) are listed in Table~4. 

\begin{table*}[H]
\begin{center}
\scriptsize
\caption{\small Best fit parameters of our radial velocity measurements. Best fit parameters are in bold face(see text for more explanations).}
\begin{tabular}{ccccc}
 &  &    &  & \\
Method  & Em. Line & $\gamma$ & K$_1$ & $\phi_{R/B}$ \\
 &  &    &  & \\
Double Gaussian  &  H$\alpha$ &  243$\pm$14  &  103$\pm$20 & 0.28$\pm$0.03\\
line flux barycenter  & H$\alpha$  & 247$\pm$14 & 125$\pm$19 & 
0.31$\pm$0.03 \\
 &  &    & &  \\
\end{tabular}
\end{center}
\end{table*}

The red spectra are time resolved, thus, radial velocity measurements of the 
H$\alpha$ emission line were obtained. 
We made use of two different methods to measure the H$\alpha$ emission line radial velocity: i) the double Gaussian fit method (Shafter 1983, Schneider \& Young 1980), and ii) the line flux barycenter. 
The best fitting sinusoidal curves for the two cases are in Table~5 and Fig.~3. 
The two methods provides best fit sinusoidal curves which agree within the uncertainties\footnote{However, we will use the double Gaussian method best fit in all our analysis.}. Measures of the line flux barycenter are expected by definition to be biased by the hot spot emission which breaks the symmetry of the accretion disk emission line. The double Gaussian fit method, when fitting the wings of the accretion disk emission lines is assumed to be free hot spot bias. Thus, the fact that the two sinusoidal curves are statistically identical possibly implies that there is no a significant hot spot contribution in the emission lines biasing the radial velocity measurements. 
However, it must be noted that the derived spectroscopic inferior conjunction is considerably offset with respect to the photometric phase zero (with an offset of $\sim$0.28, which is much larger than the uncertainty on the real eclipse minimum time in our photometric light curve). Similar phase offsets are common in other dwarf novae (see Mason et al. 2000, and references therein). Though phase offsets are expected to be large in WZ~Sge type objects (Mason et al. 2000), there is evidence for large phase offset also in systems which do not show strong hot spot contribution (e.g. Mason et al. 2001). However, OU~Vir phase offset appears exceptionally large, the typical values in the literature being in the range 0.02-0.2, in dwarf nova  systems (Mason et al. 2000). 
To further investigate the hot spot fractional contribution to the OU~Vir red light we made an H$\alpha$ trailed spectrum and back projected Doppler map (Figs.~4 and 5, respectively). 

\begin{figure}
\rotatebox{-90}{\resizebox{15pc}{!}{\includegraphics{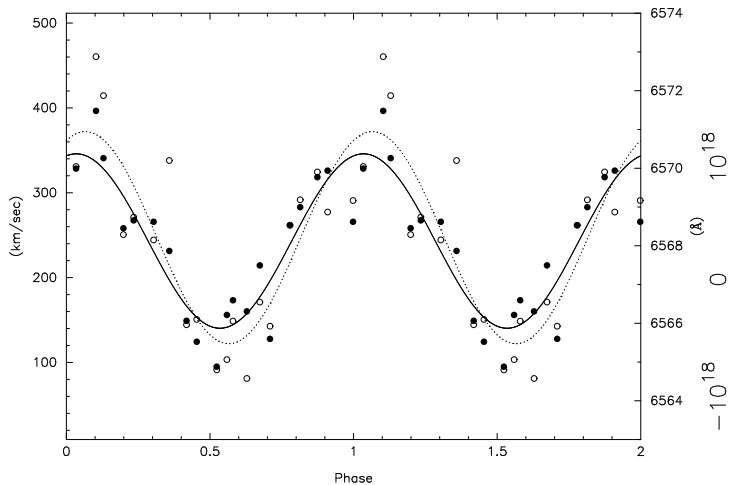}}}
\caption{The H$\alpha$ radial velocity measurements and their best fit derived through the two different methods described in the text. Solid circles and solid lines represent, respectively, the measurements and best fit determined through the double Gaussian fit method. Empty/white circles and dotted line represent, respectively, the radial velocities measured through the line flux barycenter and their best fit.}
\end{figure}

\begin{figure}
\rotatebox{-90}{\resizebox{15pc}{!}{\includegraphics{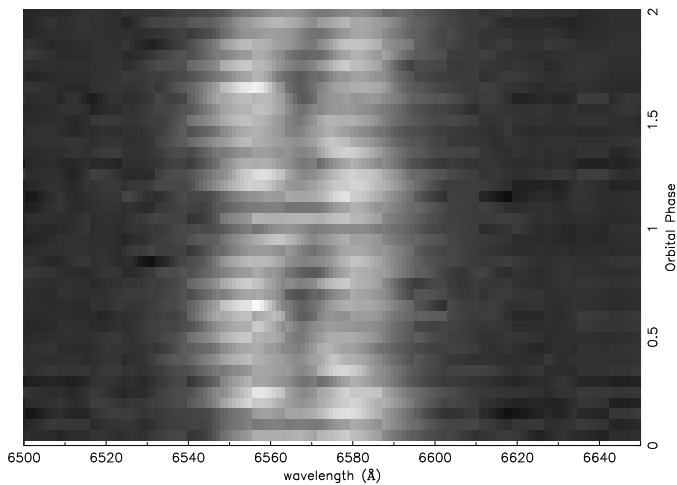}}}
\caption{\small Trailed spectrograph of the H$\alpha$ emission line. }
\end{figure}

\begin{figure}
\rotatebox{-90}{\resizebox{16pc}{!}{\includegraphics{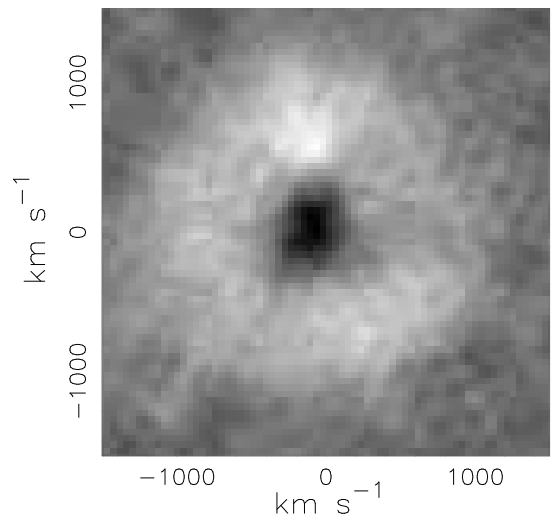}}}
\caption{H$\alpha$ back projected Doppler map of OU~Vir. 
Before back projection, spectra were v-binned and Fourier filtered (FWHM of the filtering function was set to 10. See Horne 1991 for details about Doppler mapping). No eclipse spectra were excluded due to the fact that no eclipse signature was visible in the emission line flux at phase 0 (e.g. Fig.~7) , and that large offset between photometric and spectroscopic secondary inferior conjunction (see text). The input $\gamma$ velocity we used is 243 km sec$^{-1}$ (see Table~5).}
\end{figure}

The trailed spectrum in Fig.~4 shows only marginal evidence for a S-wave component. Fig.~4 also shows that the S-wave component is possibly asymmetric moving faster from the red-to-blue than from the blue-to-red accretion disk emission peak. The times of the maximum red and blue shifts of the S-wave are not 0.5 phase apart, but occur at phases 0.22 and 0.60, respectively. Similarly, the red to blue crossing and the blue to red crossing of the S-wave seem to occur at phases 0.02 and 0.45, respectively.

The back projected Doppler map (Fig.~5) shows a quite uniform accretion disk with a stronger emission ($\sim$30\% stronger than the accretion disk), roughly at the secondary star postion (v$_X\sim-$125 km sec$^{-1}$, v$_Y\sim$600 km sec$^{-1}$). 
Interpretations of both the S-wave in Fig.~4 and the bright spot in Fig.~5 are not obvious. If real, the asymmetry of the S-wave component pointed out above, excludes it is originating from a source in circular Keplerian motion such as the secondary star. Moreover, the bright spot in the Doppler map, if interpreted as secondary star emission, implies irradiation by the hot spot (which is not emitting in the lines), rather than the white dwarf\footnote{We expect the secondary star emission to be symmetric with respect v$_X$=0 in the case it is irradiated by the white dwarf.}. 
We must not forget that the spectra are phased according to our May 1$^{st}$ V band light curve and that there is a large real offset between the spectroscopic and the photometric time for the secondary star inferior conjunction. Phasing of the spectra according to the derived time of the red to blue crossing would rotate counterclockwise the Doppler map by $\sim$103$^o$ and place the bright emission closer to the expected hot spot position\footnote{More precisely the bright emission would appear almost in the lower left quadrant, i.e. ahead of the standard hot spot position which is expected in the upper left quadrant.}. 

A straightforward conclusion is that the accretion disk emission lines are not good tracers of the primary star Keplerian motion (no matter if the hot spot is present or not!), and that the radial velocity measurements do not correspond to the white dwarf orbital motion.  

\section{Discussion and Conclusions}

We have observed asymmetric eclipses in all four optical 
band-passes for the high orbital inclination system OU~Vir. 
Comparison of the eclipse shape and depth with that observed in other 
high orbital inclination systems implies that OU~Vir is most likely 
a system for which a grazing eclipse of the hot spot occurs. The fact that we do not observe a white dwarf eclipse, constraints the inclination of the orbital plane within the range 60$^o < i < 75^o$\footnote{These limits have been computed assuming standard Roche lobe geometry for $q=0.15-0.25$ (see, e.g., Warner 1995, pg 33). The upper limit corresponds to the constraint of no white dwarf eclipse at phase 0. The lower limit is the constraint for no accretion disk eclipse at phase 0. The lower limit would be slightly smaller if assuming some hot spot size/volume.}. 
Considering that both the orbital inclination, $i$, and the two star components mass are unknown, we plot in Fig.~6 the $q$-$i$ and $q$-M$_2$ relations derived from formula (2.79) in Warner (1995), and valid for OU~Vir (i.e. K1=103 km/sec, P$_{orb}$=104.7 min). M$_2$ and $i$ were set to a constant value in the $q$-$i$ and $q$-M$_2$ relation, respectively. In Fig.~6, OU~Vir's $q$-$i$ relation is defined for $q\geq0.28$, and predicts $q$ values in the range 0.3-0.35 in the case of $60^o\leq i \leq 75^o$. Similar large $q$ and M$_2$ values can be inferred through $q$-M$_2$ relation and high orbital inclination. 
Thus, our predicted $q$ ratio is larger than typically observed in dwarf novae below the period gap (e.g. OY~Car 0.102, Z~Cha 0.15, and HT~Cas 0.15, Warner 1995 and references therein), though in rough agreement with computation by Webbink (Warner 1995). 

\begin{figure}
\rotatebox{-90}{\resizebox{15pc}{!}{\includegraphics{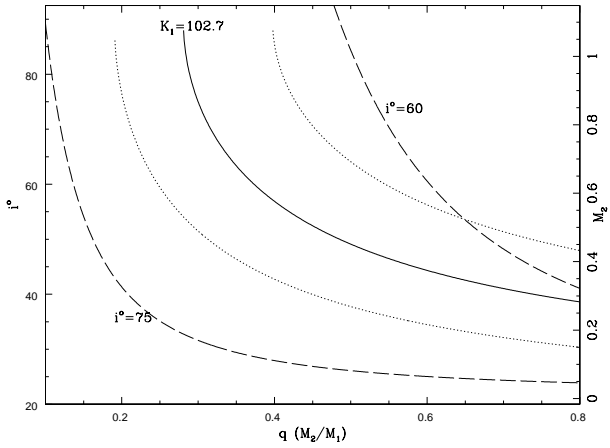}}}
\caption{The $q$-$i$ and $q$-M$_2$ relations as derived solving for formula (2.79) in Warner (1995). Values for $i$, in deg, are on the left side y-axis; value for M$_2$, in solar units, are on the right side y-axis. The solid line represents the $q$-$i$ relation valid for K$_1$=103 km/sec, P$_{orb}=104.7$ min, and M$_2$=0.17 M$_\odot$ (a secondary star mass of 0.17M$_\odot$ is predicted by the P$_{orb}$-M$_2$ relation in Howell et al. 2001, for systems having the same orbital period of OU~Vir and before the orbital period minimum). The two dotted lines represent the $\pm 1\sigma$ error for the Keplerian velocity of K$_1$=103 km/sec. Dashed lines represent the $q$-M$_2$ relation for constant values of the orbital inclination, $i=60^o$ and $i=75^o$. }
\end{figure}

The derived eclipse colors (see Sect.~3.1) possibly imply a contribution by the secondary star. Signature of the secondary star appears also in the red-ward rising slope  beyond 7200\AA \ (e.g. Fig.~2). Unfortunately the spectral coverage and the lack of secondary absorption features do not allow us to set constraints on the spectral type of the secondary star. 

Analysis of the light curve features (see Table~3) allows us to estimate 
a hot spot contribution of $\sim$40\% in the V band and 
26\%, 17\%, and 10\% in the B, R and I bands, respectively. 
 Similar contributions of 20\%, 58\%, and 9\% (in B, V, and R 
respectively) were found in OY~Car by Schoembs et al. (1987). 
WZ~Sge, which has weak continuum emission but a strong hot spot 
emission line contribution, has fractional hot spot contribution of only 
8\%, 6\%, and 12\% in the B, V, and R bands\footnote{The WZ~Sge hot 
spot fractional contributions have been computed using high speed 
multi-color photometry obtained on August 16$^{th}$ 1991 by W. Skidmore
(private communication).}, respectively. 

It is more difficult to compute the hot spot fractional contribution in the emission lines.  In first place, the two components of accretion disk and hot spot cannot be deconvolved in Doppler maps. Secondly, in the case of OU~Vir (see previous section) there is not obvious {\it a priori} identification of the bright emission in the Doppler maps and/or the S-wave in the trailed spectrograph, due to mismatching photometric and spectroscopic phase 0.  

The accretion disk peak intensity ratio V/R is often used as a diagnostic to point out asymmetries in the accretion disk itself. In particular the V/R ratio is expected to be modulated by the hot spot emission and show values greater than 1 when the hot spot is on the approaching side of the disk gas, and values smaller than 1 when on the disk side which is moving away from the observer. The maximum value should occur around phase 0.66 for an hot spot in the standard position predicted by the stream trajectory (Gilliland 1982), or in the phase range 0.1-0.3 in the case of a system showing the reverse hot spot phenomenon (Mennickent 1994 and references therein). However, we often measured a smaller intensity value in the disk peak which is blended with the hot spot, contrary to the expectations. The plot of V/R ratio vs. orbital phase shows mostly random scatter and no hot spot modulation.  

\begin{figure}
\rotatebox{-90}{\resizebox{15pc}{!}{\includegraphics{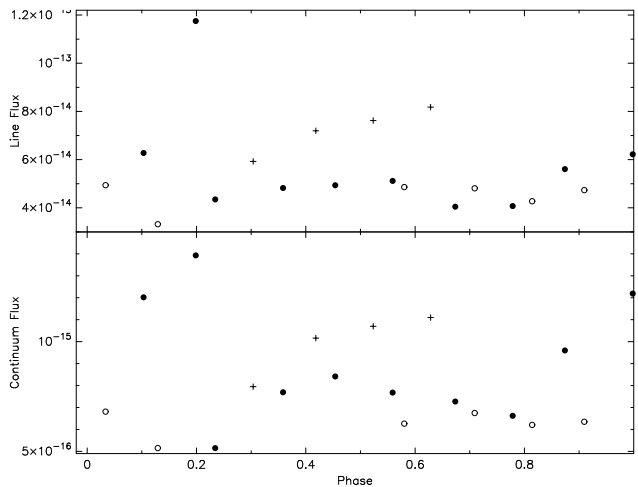}}}
\caption{Emission line (top) and continuum flux (bottom) vs. orbital phase. Both the continuum and the line flux were measured by the {\it ee} command in the IRAF task  {\it splot}. The continuum flux results from the average of the two points marked by the cursor position; the line flux is the sum of the counts within the two cursor position after continuum subtraction. Different symbols are used to show different orbital cycles. }
\end{figure}

Similarly, a plot of the line and underlying continuum flux vs. phase (Fig.~7) show large scatter and poor modulation. Both light curves appear strongly affected by accretion system intrinsic variation on the timescale of an orbit or less\footnote{We are inclined to interpret the dramatic increase in flux from phase $\sim$0. to 0.1 as a flare episode.}.

The blue spectrum was used to compute the Balmer decrement of H$\delta$/H$\beta$, to be compared with theoretical models of accretion disks. The derived value of 0.75 (frequency units) is consistent with the intensity ration predicted by Williams (1980) in optically thick emission lines on low mass transfer rate accretion disks ($\sim 10^{-12}$ M$_\odot$ yr$^{-1}$). The H$\alpha$/H$\beta$ flux ratio was not computed due to the mismatching continuum of the blue and the red spectra (see Fig.~2). The measured H$\delta$/H$\beta$ ratio is of 0.55 but there are no models predicting such a line ratio to compare with.  

In summary, BVRI band time resolved photometry of OU~Vir in quiescence shows hot spot modulated light curve and deep eclipse as typical in high inclination SU UMa stars (e.g. OY~Car Wood et al. 1989, Z~Cha Wood et al. 1986, van Amerongen et al. 1990). However, the eclipse depth suggests that only a grazing eclipse of the hot spot is actually occurring in OU~Vir, similarly to WZ~Sge (e.g. Skidmore 1998) and V893~Sco (Bruch et al. 2000). 
Our photometry also detected cycle to cycle random brightness variations up to 0.25 mag, which -though smaller- are possibly similar to the cycle to cycle variation reported by Bruch et al. (2000) for V893~Sco.   
Random brightness variation are also observed in the OU~Vir H$\alpha$ emission line. The line flux light curve together with the line profile analysis suggests that these brightness variations occur in the whole disk and are not restricted to the impact region. No matter how we interpret the S-wave and the bright emission in the Doppler map, the OU~Vir hot spot results suggest an optically thick region which does not produce strong emission lines.  
At last, both the red spectra continuum and the eclipse colors seem to indicate evidence of the secondary star, which has only rarely been detected in short orbital period dwarf novae (Z~Cha, Bailey et al. 1981).

All these characteristics make OU~Vir an extremely good target for the solution of the binary geometry and the modeling of the accretion system (i.e. disk plus hot spot). However, those results are presently made difficult by the observed strong flux variations and both the time and spectral resolving limits of our data sets. Fast speed photometry would be useful to derive the accretion disk flux distribution through eclipse analysis, and to interpret its brightness variation, while both IR spectroscopy and/or photometry should be pursued to possibly detect the secondary star, derive its temperature, Keplerian velocity, K$_2$, and mass. 

\begin{acknowledgements}
Observations reported here were obtained at the MMT Observatory, a joint
facility of the University of Arizona and the Smithsonian Institution.
This research was partially supported by NSF grant AST 98-10770 to SBH.
\end{acknowledgements}

\end{document}